\documentclass[3p,twocolumn,authoryear,12pt]{elsarticle}

\usepackage{epsfig}
\usepackage{graphicx}
\usepackage{float}

\usepackage{amssymb}
\usepackage{amsmath}
\usepackage{algorithmic}
\usepackage{array}
\usepackage{natbib}
\usepackage{hyperref}

\journal{Advances in Space Research}

\begin{document}

\begin{frontmatter}



\title{Advances in Deep Space Exploration via Simulators \& Deep Learning}


\author{James Bird\corref{cor1}%
 \fnref{fn1}}
\ead{james\_bird@ucsb.edu}


\author{Linda Petzold\fnref{fn1}}
\ead{petzold@ucsb.edu}

\author{Philip Lubin\fnref{fn2}}
\ead{lubin@ucsb.edu}

\author{Julia Deacon\fnref{fn3}}
\ead{jcdeacon@alumni.caltech.edu}

\cortext[cor1]{Corresponding author}
\fntext[fn1]{Department of Computer Science, University of California Santa Barbara, 2104 Harold Frank Hall, Santa Barbara, CA 93106-9530}
\fntext[fn2]{Department of Physics, Broida Hall, University of California Santa Barbara, Santa Barbara, CA 93106-9530}
\fntext[fn3]{Department of Computing and Mathematical Sciences, California Institute of Technology, 1200 E. California Blvd., MC 305-16, Pasadena, California 91125-2100}

\begin{abstract}

The NASA Starlight and Breakthrough Starshot programs conceptualizes fast interstellar travel via small relativistic spacecraft that are propelled by directed energy. This process is radically different from traditional space travel and trades large and slow spacecraft for small, fast, inexpensive, and fragile ones. The main goal of these wafer satellites is to gather useful images during their deep space journey. We introduce and solve some of the main problems that accompany this concept. First, we need an object detection system that can detect planets that we have never seen before, some containing features that we may not even know exist in the universe. Second, once we have images of exoplanets, we need a way to take these images and rank them by importance. Equipment fails and data rates are slow, thus we need a method to ensure that the most important images to humankind are the ones that are prioritized for data transfer. Finally, the energy on board is minimal and must be conserved and used sparingly. No exoplanet images should be missed, but using energy erroneously would be detrimental. We introduce simulator-based methods that leverage artificial intelligence, mostly in the form of computer vision, in order to solve all three of these issues. Our results confirm that simulators provide an extremely rich training environment that surpasses that of real images, and can be used to train models on features that have yet to be observed by humans. We also show that the immersive and adaptable environment provided by the simulator, combined with deep learning, lets us navigate and save energy in an otherwise implausible way. 
\end{abstract}


\end{frontmatter}

\parindent=0.5 cm


\section{Introduction}

Space travel, up until recently, was constrained by chemical propulsion, large spacecraft, and therefore, relatively slow speeds. Since the main objective has been exploration of our solar system, these methods were sufficient. In contrast, the recent Starlight program \citep{KulKarniRelativistic2017} has introduced methods for deep space travel that utilize small discs, which travel at approximately one-fourth of the speed of light via directed energy.

Alongside the prospect of fast deep space travel comes many new challenges. The normal model for space travel includes spacecraft capable of housing instruments, propulsion and navigational equipment, telescopes, energy banks, and much more. Since the Starlight program will be utilizing small wafersats that are approximately the size of a coffee can lid, all of these features need to be reworked or discarded.

Besides physical constraints, this new model of space travel introduces feasibility constraints as well. The star of interest is beyond four light-years away, meaning that transmission of data and response command transmissions are a combined eight years or more. Thus, the wafersats need to be able to make decisions without human intervention, and for that, artificial intelligence (AI) is paramount.

The major hurdles that we will discuss are those concerning computer vision via planetary detection, data and storage blockages via novelty detection and ranking, and energy management via combining simulator features with subtraction-algorithm-fed computer vision. For all of these issues, taking advantage of a universe simulator will introduce solutions that were otherwise ineffective or impossible to find. 

\subsection{Previous Work}

The effectiveness of machine learning, specifically deep learning via TensorFlow and cuDNN, has been indisputably demonstrated in the last decade (\cite{Abadi2016}, \cite{ChetlurcuDNN2014}, \cite{Canziani2016}). The fight over the best model and the most accurate results, especially between the most popular models like ResNets, DenseNets (\cite{HuangDensely2018}), Inception(\cite{SzegedyRethinking2015}), Masks (\cite{HeMask2018}), and models that combine some of these together(\cite{Szegedy2016}), is one that has produced a plethora of potent options to choose from. Models that are more accurate than human beings at doing extremely difficult tasks are still being discovered (\cite{RajpurkarChexNet2017}).

The areas of deep learning and astronomy have come together in recent years (\cite{RuffioBayesian2018}, \cite{MoradOnOrbit2018}, \cite{Schaefer2018}, \cite{PearsonSearching2017}), mostly in the form of light curves (\cite{ShallueIdentifying2017}, \cite{ZuckerShallow2018}, \cite{CarrascoDeepLearning2018}). The results and general concepts promote a healthy symbiosis between deep learning and the problems that arise in astronomy. Yet, the processes are carried out from Earth, not space, and do not address real images, two big issues that create a gap in comparability. 

Outside of astronomy, simulators have been used to train data in specific instances where the benefits outweigh the drawbacks. \cite{SmythGame2018} outlines some major drawbacks, namely that the process takes a lot of time and knowledge, as well as a note that simulator-based training may not generalize well to real images. Alongside those concerns, \cite{McDuff2018} begins with a common issue in machine learning models, which is that training data sets are often biased. This bias arises when there are minorities in the training set, which in turn produces poor results when the model is asked to evaluate a similar entity in the population. These issues are handled throughout this paper and are shown to not be an issue with the specific problem at hand. 

Simulators also introduce a lot of benefits. One large one, also seen in \cite{ConnorApplying2018}, is that "the small catalogue of real events is probably not yet a representative sample of the underlying .. population, nor is it big enough to build a meaningful training set for machine learning, deep or otherwise." An important theme throughout this paper, and an extremely useful aspect of simulators, is that they provide an untold amount of training data, assuming that one can create realistic simulations.

\subsection{Unsupervised Learning for Planetary Detection}

The intuition behind object detection, in particular planetary detection, might point toward an unsupervised learning technique. After all, one might reasonably think that detecting a nearby planet after months of traveling through deep space would be easy. We test this idea using an unsupervised technique called a \textit{Grow When Required (GWR) Network} \citep{gwr}.  

\subsubsection{GWR Setup}

Using the worldwide telescope, we generated a 9,000 frame series of solar system images.  It begins with Neptune, then it explores Mercury, the Sun, and finally Mars. The majority of images contain only background stars.

The images were down-scaled to a 320x180 resolution in order to improve computational speed.  For learning, they were decomposed into red, green, and blue channels and vectors were constructed of length $320 \times 180 \times 3 = 172,800$.

Our challenge is to label each image as novel or regular.  That is, we wish to generate a classification $n$ s.t. for each input $x$, $n(x) \in \{0, 1\}$, where a 0 indicates regularity and a 1 indicates novelty.  The algorithm should hopefully yield a large number of $1$'s when a planet or the Sun is clearly in view and should yield very few $1$'s when the image is mostly distant stars.
\subsubsection{GWR Algorithm}

Define $A$ as the set of nodes in our network and $C$ as the set of edges between these nodes.  We denote our inputs as $\boldsymbol \xi$ and the weight vector for any node $n$ as $\boldsymbol{w_n}$.  Each node $n$ has a habituation $h_n$ which represents how familiar that node is to the system.
\begin{enumerate}
	\item Initialize two nodes that represent two random samples from the dataset.  We set their habituations each to 1.  The set of edges between nodes begins empty.
	\item Iterate through the dataset.  For each data sample $\xi$:
	\begin{enumerate}
		\item For each node $i$ in the network, calculate its distance from $\xi$, which is $||\xi -{w_i}||$.
		\item Find the node $s$ with the smallest distance and the node $t$ with the second-smallest distance.
		\item Add an edge between $s$ and $t$ if it does not already exist.
		\item Calculate the activity $a = \exp(-\sum_j ({\xi}[j] - {w_s}[j] )^2/C)$, where $C$ = 29,203,200 was chosen to prevent an integer overflow.  There are 172,800 fields in each data vector, and since the average of the quantities in each vector is close to 13, and $13^2 = 169$, we divide by $172,800 \times 169 = 29,203,200$.
		\item If $a < a_T$ and $s$'s habituation $h_s < h_T$ (where $a_T$ is some insertion threshold and $h_T$ is some habituation threshold), then add a new node $r$.  Set ${w_r} = \frac{{w_s} + \xi}{2}$.  Insert edges between $s$ and $r$ and $r$ and $t$ and remove the edge between $s$ and $t$.
		\item Otherwise, update the weight and habituation of $s$ as follows: $\delta {w_s} = \epsilon_b \times (\xi - {w_s})$ and $\delta h_s = \tau_b \times 1.05 \times (1 - h_s) - \tau_b$, where $\epsilon_b$ and $\tau_b$ are parameters.  Next, update the weight and habituation of $s$'s neighbors $i$ as follows: $\delta {w_i} = \epsilon_n \times (\xi - {w_i})$ and $\delta h_i = \tau_n \times 1.05 \times (1 - h_i) - \tau_n$, where $\epsilon_n$ and $\tau_n$ are other parameters.
		\item Remove any nodes without any neighbors.
	\end{enumerate}
	Our chosen values are: $a_T = 0.7, h_T = 0.1, \tau_b = 0.3, \tau_n = 0.1, \epsilon_b = 0.1, $ and $ \epsilon_n = 0.01$.
\end{enumerate}

\subsubsection{Results}
\begin{figure}[ht!]
	\begin{center}
		\includegraphics[width=\linewidth]{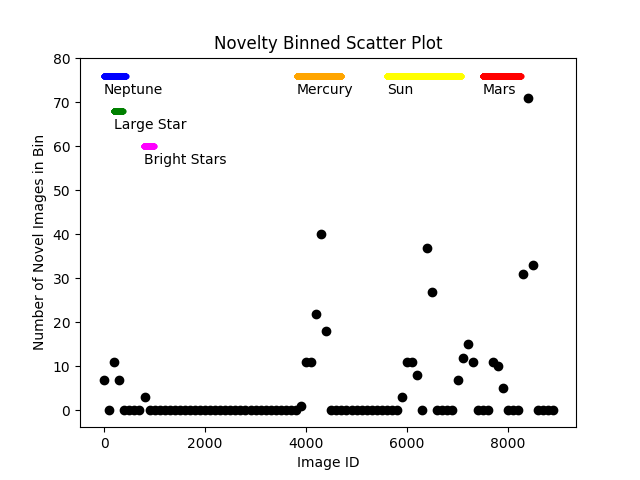}
	\end{center}
	\caption{A scatter plot of the detected novelty of the data.}
	\label{fig:binned_scatter}
\end{figure}

\begin{figure}[ht!]
	\begin{center}
		\includegraphics[width=\linewidth]{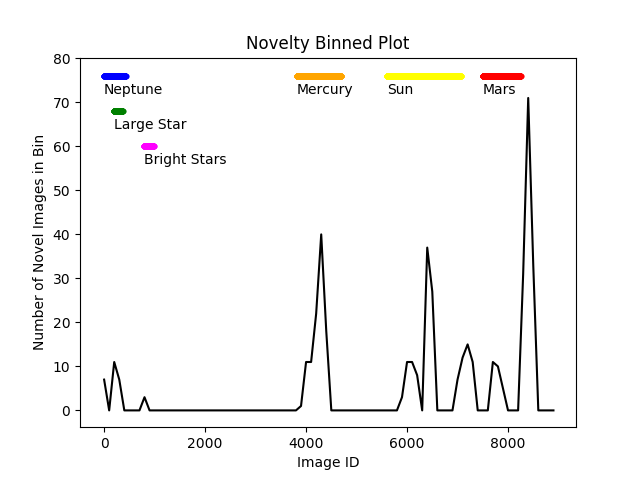}
	\end{center}
	\caption{A binned scatter plot of the novelty of the data.  Image ranges that are salient to the human eye are labeled on the plot.}
	\label{fig:bin}
\end{figure}

Figure \ref{fig:binned_scatter} is a scatter plot that was generated to visualize the novelty detected from the data.  The $x$-axis is the id of each picture, and the $y$-axis is the number of novel images that were detected in each bin of 100 images. Figure \ref{fig:bin} is a continuous representation of the same concept.

Moving along the Image ID axis, we see that novelty was detected in clumps around 0-500, 900, 4000-4500, 6000-6500, 7000-7500, 7700-8000, and 8200-8500.
We observed that novelty was detected first on Neptune, then again on some particularly bright stars.  No novelty is detected during the long period of only stars.  Next we see increased novelty detection when Mercury is plainly in view, and then when the sun appears, and finally when we zoom into Mars.  

We notice that Neptune's collection of novelty is roughly one quarter the size of the other three celestial objects that come into view.  We also notice a huge spike around image 8300.  This is very interesting because there are no large celestial objects in view at this time.

\subsubsection{GWR Discussion}

A deep space exploration mission would come with many challenging objectives. A small but connected subset of those would involve detecting objects, deciding whether they are important, extracting key features that we would want to study or observe, and prioritizing their information retrieval. 

\textit{GWR} wouldn't be able to decide importance, extract features, or prioritize information retrieval, yet if it could detect novel objects in deep space, this would be useful. We can see from Figure \ref{fig:binned_scatter} and Figure \ref{fig:bin} that the detection is inconsistent and unreliable. Neptune is almost completely missed and the three smaller peaks at the Sun, as seen in Figure \ref{fig:bin}, are larger than Neptune. The largest peak of all happens while Mars is minuscule and essentially not in view. 

Although \textit{GWR} had high novelty detection peaks while passing by Mercury and the Sun, it failed to correctly activate at Mars or Neptune. These observations, paired with its inability to do anything further with the data, introduces a need for a more advanced model that can achieve all of the above objectives. 

\subsection{Object Detection vs. Novelty Detection}

Throughout this paper, our main goals will constantly be alluding to object detection and novelty detection. In a general computer science setting, object detection is used to identify something in an image that has already been trained via some algorithm. For example, we may feed thousands of images of human beings into a YOLO algorithm, and then once it is trained, we can walk the streets of New York and see if our algorithm can identify human beings. In this setting, identifying a human being is a success, and not identifying a car or stop sign as a human being would also be a success. Yet, identifying anything non-human as a human being would be a failure. The accuracy of a model, which is mathematically computed per identification, can be used as a measure of how sure the algorithm is that the object being identified is the correct type. In this paper, we will delve into why this is difficult for our specific scenario, and we will test whether this can benefit severely from the use of simulators.

On the other hand, novelty detection is used to attempt to identify something that has never been seen before. One powerful example is self-driving cars being able to see traffic signs that are unique to a certain country, and therefore have never been seen or used during the training process \citep{template}. In this example, the self-driving car algorithm has never seen this specific sign before, and so identifying it without any training data is very difficult. In our paper, unseen planetary features are analogous to the unseen traffic sign in the example, and we delve into methods of solving this via simulators.


\section{The Simulator}
Although there are quite a few universe simulators available today. Here, we utilized SpaceEngine (SpaceEngine.org) for its realism, expansive set of options and customizations, and unique informational tools. 

\subsection{Simulator Features}
One of the best features of the simulator is its extremely realistic rendering capabilities. In combination with a 3840x2160 4K monitor and GTX 1080 Ti, the simulator produces extremely detailed and realistic images. 
\begin{figure}[ht!]
  \includegraphics[width=\linewidth]{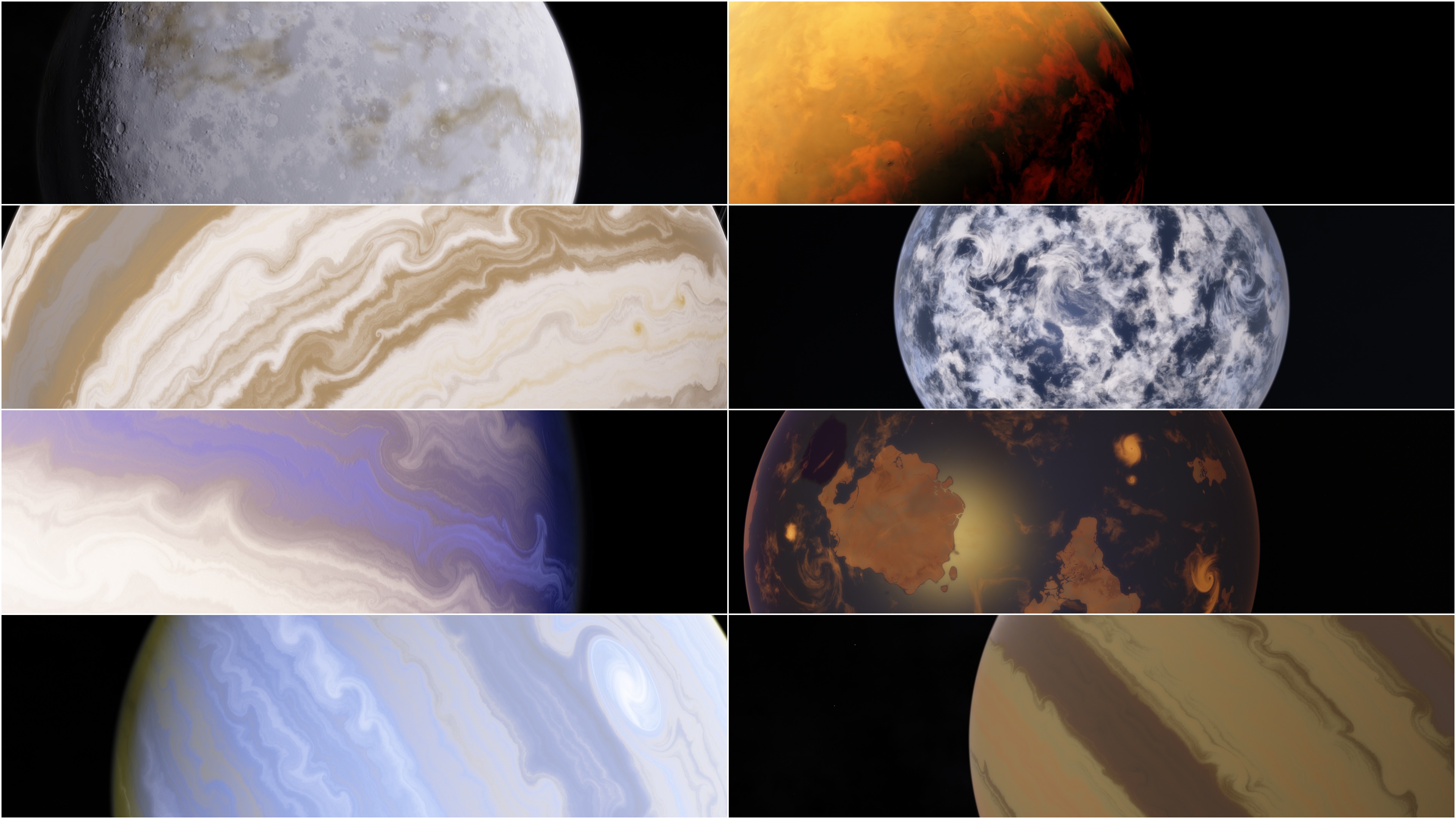}
  \caption{Examples of 3D-rendered randomly generated exoplanets}
  \label{fig:CollageExos}
\end{figure}

The simulator also includes the ability to edit any planet, so that instantly rendering an exoplanet with a very particular feature set is simple. Alongside image features are astronomical features, which are tracked and shown for every body in the universe. Some of these features are type, class, orbital period and mass, but most importantly, distance. 

\subsection{Simulator Options}
Graphical options in the simulator are abundant, which allow for complete control of the simulated universe. In general, the feature set should be optimally set for realism while traveling through space, but the ability to tweak these options speaks to greater breadth for learning and adapting to unique situations that may arise in space. For example, the image of an exoplanet while traveling at one-fourth of the speed of light with a nebula in the background is a completely new concept. Yet, two features in the simulator may be able to deal with that combination. First, the ability to toggle lens flares will provide  the AI with training images that both contain and lack lens flares. Second, a feature called overbright can drastically adjust how bright the background stars and nebula appear. Training on images that embrace the entire spectrum of overbright will allow this machine to deal with novelty detection in a very advanced manner. Having a plethora of options to enable and tweak can introduce a much larger set of training images and will let the AI absorb more information before embarking into deep space. 

\begin{figure}[ht!]
  \includegraphics[width=\linewidth]{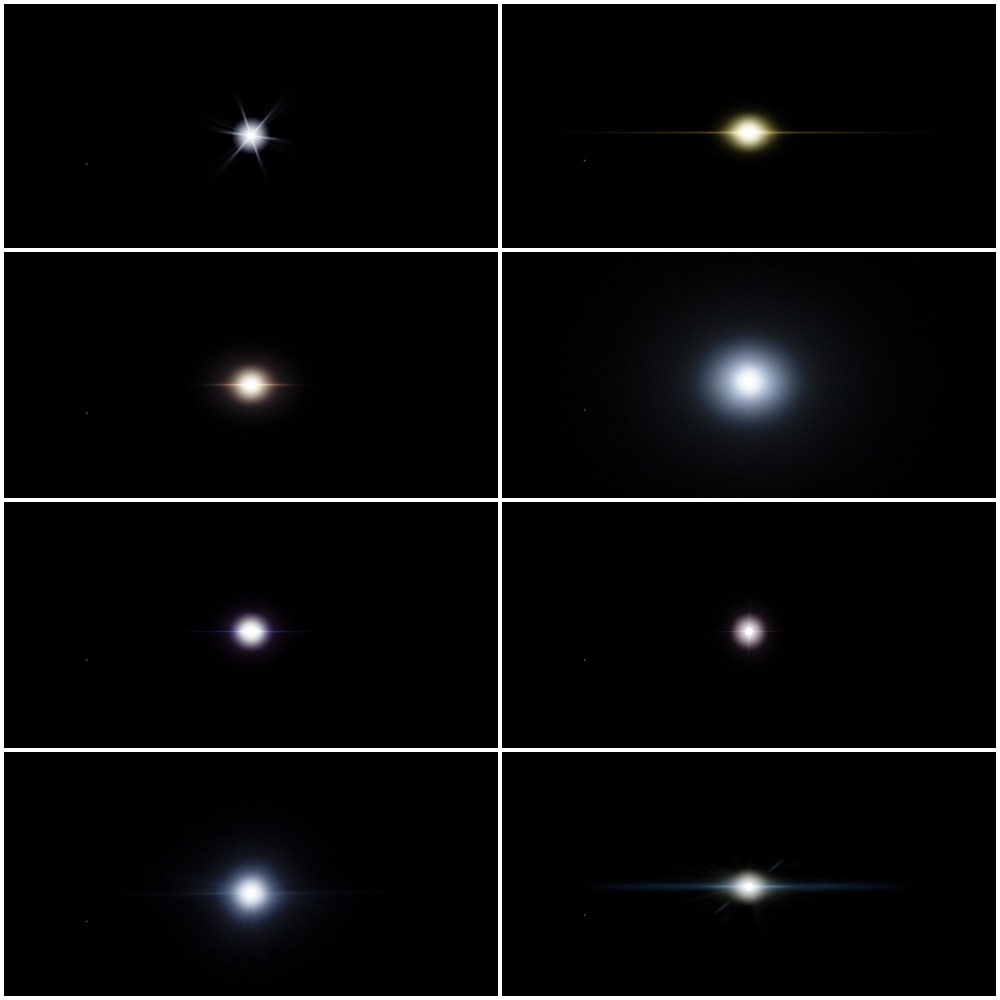}
  \caption{Eight lens options applied to the same star}
  \label{fig:StarLens}
\end{figure}

Some other important options, besides those that deal with graphics and rendering, are diffraction spike intricacy and size, lens effect on stars, and planetary shine. Altering all of these settings and training on the resulting images enables the capture of more information.

\subsection{Overall Simulator Importance}

In this paper, We consider three main areas of deep space travel that can be drastically improved with the use of a simulator. 

First, computer vision is an extremely useful tool for detecting objects and making decisions based on what is seen. The training process consists of tagging images and providing a label for each tag, feeding those images and tags into a model, and having that model learn the associations. The model can then be given images, and based on how successfully it was trained, it may be able to identify  parts of the image.

Since we have never photographed exoplanets in detail, training a model using real images is not feasible. Therefore, we rely on training using images of planets that we have photographed, which would be those in our own solar system. Yet, detailed photographs of planets are not very abundant and would only teach the model to look for those specific features. In realizing that this would not be sufficient, we may move toward novelty detection, a branch of computer vision that tries to classify data that deviates from the data used during training. Co-domain embedding \citep{template} has proven useful in some situations, such as those where a template design would resemble a real image almost exactly, but planetary features do not translate well to use in novelty detection. This is because planetary features, such as atmospheric patterns, are extremely unique. 

Simulators can provide very detailed and randomly generated images of planets that obey universal physical laws. Therefore, we will be able to generate countless images of planets that resemble real images of possible exoplanets. Training on these images and features, the model will learn an exorbitant amount of information. While traveling through space and faced with an image of a real exoplanet, the model will now have a much broader knowledge base.

Second, we introduce the notion of novelty ranking. A major hurdle in deep space wafersat travel is data storage and transmission. On-board memory is limited by physical constraints of the wafersat and astrophysical exposure, while transmitting data from a wafersat to a communications hub would be slow and dependent on energy reserves. 

A system that can deal with this issue is one that prioritizes the most important on-board data and sends that first. This not only ensures that the critical images are sent in descending order of importance in case of some malfunction, but that the most relevant data is quickly known for the next wafersats in line. 

With the overarching goal being the identification and transmission of the most important data, novelty ranking will quantify the on-board images based on importance. Simulators will provide the breadth of planetary features that are needed to find out what \textit{importance} means, as perceived by humans, and then this information can be applied to software.

Third and last, sending a small disc into deep space means that on-board energy reserves will be very small. Yet, the objective of detecting and imaging astronomical bodies while traveling must still be met.



\section{Simulator for Planetary Detection}

Our main objective here is to identify novel planets while traveling through deep space. In order to do so, and for subsequent sections, we will require a basic conceptual understanding of object detection in order to logically progress. We should point out that the main backbone of object detection, through a few core processes, is the same as that used by humans when they naturally process information and identify objects. We will discuss these fundamental core processes. 

First, the object that we will try to have the machine identify should be seen beforehand in order to train the model. Unsupervised techniques have their uses and do not require this training process, but we will only deal with supervised learning models from here on out. Mainly, this is done because planetary detection is the simplest task, so we need a model that can adapt afterwards in order to successfully identify planetary features and rank novelty.

Second, the machine will learn these trained models better with more images. Of course there are exceptions here, such as feeding poor images or images that do not match the objects category. We will test this concept thoroughly while we also test the importance of simulated images.

\subsection{Setup}

Here we will discuss the details of our model, our hypothesis, and how we will go about testing the importance of simulator images. Our main goals when choosing a model are finding one that has high accuracy, low to medium computation time, and has been tested to be a reliable model. Because of this, no new models that haven't had time to be tested thoroughly throughout the computer vision community will be used. Also, the ideal model will sacrifice computation time for accuracy, if needed. Since we will be testing different combinations of image sets with the same model, our model choice will be a control and therefore we will again stress the importance of reliability and accuracy over computation time or creation date. 

Our hypothesis stems from our second core process and states that simulator images will not decrease accuracy for planetary detection and planetary features. These two processes, the detection of planets while traveling through space, as well as the detection and recognition of features on those planets, are the inspiration for the two main experiments that are set up.
Currently, our collection of useful astronomical images is very limited. Therefore, using only real images of planets would limit us to those found in our solar system. Also, planetary features would suffer since our solar system contains very few features out of the set of total planetary feature combinations. 

The first experiment will test the validity of simulator images in general. It is set up in three different stages using the same object detection framework and always testing on the set of real images of Jupiter. First, we will train on real images of every planet in our solar system except for Jupiter. These images will be collected from NASA image repositories and will not include composite images, artist renditions, or any other variations except for true unaltered images. Second, we will train on only simulator images with the goal of solidifying whether simulator images alone are useful in detecting real planets. Third, we will combine the first and second training sets, comprised of simulator images and all real images (excluding Jupiter),to determine whether simulator images and real images together provide the best of both worlds. 

The second experiment will introduce and test an extremely important feature of using simulator images - the ability to detect novel planetary features, i.e. those which have never been seen in any real images. Since simulators can be programmed to emulate real physics, the outcome can give us an extremely large number of realistic looking planets with features that have never been observed. In order to proceed, we need to use a planetary feature that exists in our solar system so that we can train using simulator images and test using real images. Planetary rings have a solid theoretical foundation and would easily appear in any physics-based simulator, while also being present around Saturn. Rings are also fairly complex, as they contain extremely unique striation patterns, can look wildly different depending on the viewing angle, and can even co-exist with other rings around the same host planet. 

\subsection{The Model}

When using deep learning models for a specific purpose, it is imperative that a model is chosen that optimizes what it can while prioritizing what it must. For example, an object detection model that may be implemented on a smart phone for real-time detection of human faces might prioritize speed and give up a small amount of accuracy. 

For our purposes, accuracy is of the utmost importance, while operation count is also of some importance. In deep space, we have plenty of time to do calculations, but we also have very little energy. Therefore, reaching maximum accuracy with a small amount of operations is the ideal scenario.

\begin{figure}[ht!]
	\begin{center}
		\includegraphics[width=\linewidth]{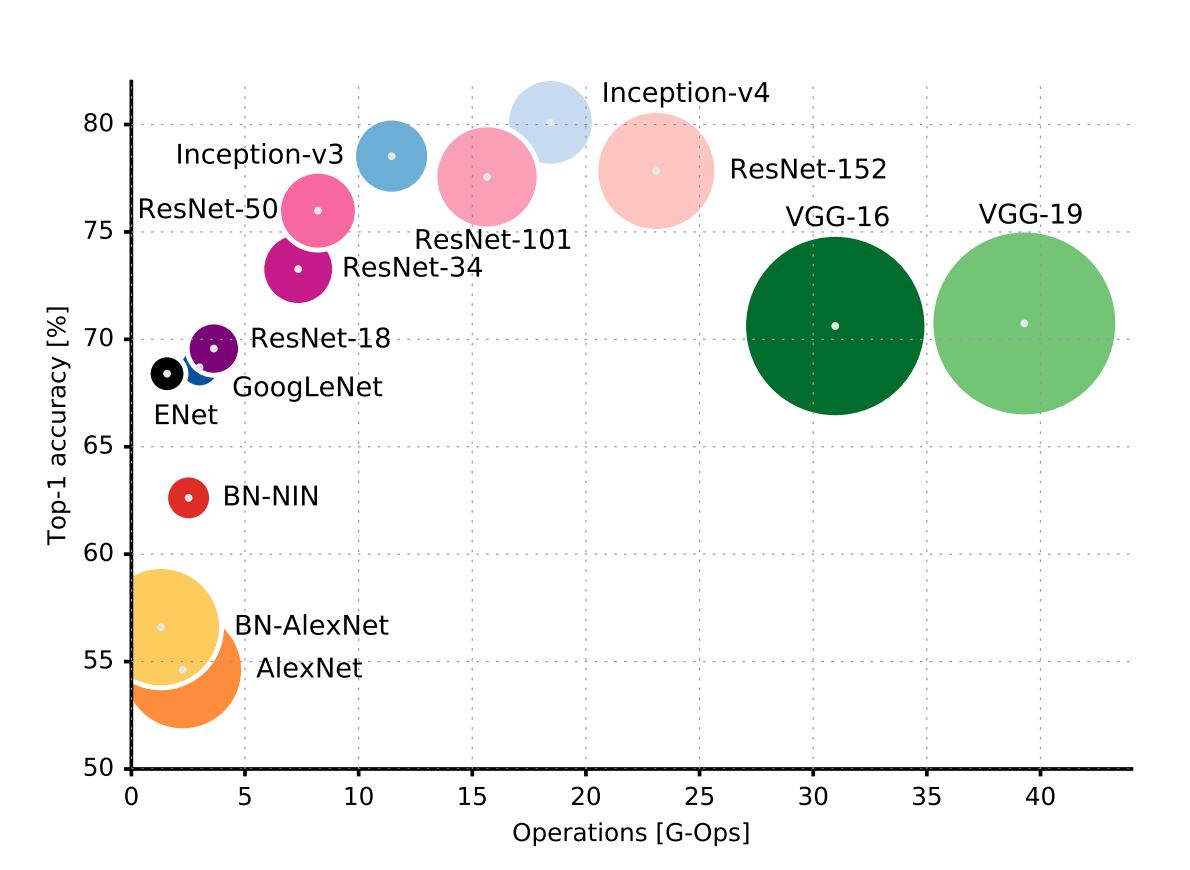}
	\end{center}
	\caption{Results shown in Canziani et al.(2016) that compare model accuracy vs. operation count during the ImageNet challenge.}
	\label{fig:Canziani}
\end{figure}

As we can see from Figure \ref{fig:Canziani}, originally presented in \cite{Canziani2016} whereas the authors compared many models for practical applications like this one, there are few models that fit into the optimal space of accuracy and operations. The main choice was ResNet architecture vs. Inception architecture. The accuracy and operations for both models are almost identical, yet the residual neural network (ResNet) architecture provides a shortcut in case the training phase introduces the vanishing gradient problem \citep{HeResNet}. Along with this feature, ResNet is a very established model in many domains, and for these reasons, will be our model of choice going forward.

\subsection{Experiment \#1 - Planetary Detection}

In this experiment, we tested the theory that simulator images could be used to train a model that could then detect real objects, and in particular, planets. Our hypothesis is that simulator images are at least as good as real images in terms of information gain during training. Although simulator images can be produced in bulk, the idea was to test the theory using similar image count in order to avoid any bias. The table below shows the number of images used for each model and for their testing phase.

\begin{table}[ht!]
\centering
 \begin{tabular}{||c c c||} 
 \hline
 Real & Sim & Real+Sim \\ [0.5ex] 
 \hline\hline
 120 & 122 & 242 \\ [1ex] 
 \hline
 \end{tabular}
\end{table}

For each of the three separate runs, the models reached a minimum of 60,000 iterations in order to ensure ample training time and accuracy. Concerning the testing images, the images were broken down into 2 sections. The first section was comprised of independent images taken at differing angles. The second section was the exact same frame of reference, including angle and distance, but included a time-lapsed series of images.

The results of the testing phase produced a model accuracy, which is a mathematical score given by the model as to how surely it has identified an object that it has previously seen during the training phase. The table below shows the final accuracy results.

\vspace{5mm}

\begin{center}
 \begin{tabular}{||c c c c||}  
 \hline
Section & Real & Sim & Real+Sim \\ [0.5ex] 
 \hline\hline
 1 & \textbf{99.875} & 99.375 & \textbf{99.875} \\ 
 \hline
 2 & 98.889 & \textbf{100} & \textbf{100} \\
 \hline
 Total & 99.353 & 99.706 & \textbf{99.941} \\
 \hline
\end{tabular}
\end{center}

\vspace{5mm}

The results show quite a few compelling results right off the bat. The most direct one being that \textit{Real+Sim} has achieved equal or better results than \textit{Real} or \textit{Sim} alone did in all categories. Besides \textit{Real+Sim}, we can also say something about \textit{Real} vs. \textit{Sim}. Although \textit{Real} achieved slightly better accuracy in Section 1, \textit{Sim} not only achieved better accuracy in Section 2, the total accuracy of \textit{Sim} was also higher and \textit{Sim} contained at least one section that had perfect accuracy. 

Our initial hypothesis was that using simulator images would be as good as real images. Different models and training image sets will always produce different results, but considering total accuracy with our two sections, \textit{Sim} produced equal or better results when compared to \textit{Real}. 

\subsection{Experiment \#2 - Novel Features}

In order to be able to show the importance of using a simulator for novel planetary features, we use the results from Experiment \#1 as proof of concept. Those show that \textit{Sim} images do not decrease accuracy on real testing images, with the added benefit of being able to mass produce them and customize feature information in each image. 

With this in mind, Experiment \#2 will gather simulator images of ringed planets and train a new model with the same framework as Experiment \#1. We will then test novel feature detection on real images of Saturn. The machine will have never seen any real image and will have never been exposed to prior knowledge of Saturn or our solar system at all. This experiment is, in theory, identical to training a model with simulator images on Earth and sending it out into deep space in order to identify novel, never-before-seen features found on real exoplanets and in real images. 

One of the main benefits of simulator images can be observed here. Even a planetary feature that we can observe will be found once, or perhaps a few times at best. Therefore, we have limited variability to work with in terms of ring structure, width, pattern, count, etc.. Yet, if this experiment produces promising results, we can simply build a physics-based simulator that generates planets, filter by the presence of rings, capture an image, and repeat the process any amount of times. From Experiment \#1, we know that training on these simulator images will net us generally equivalent information gain when compared to real images of ringed planets. Since our simulator is physics-based, it should produce many features that we have not even seen before, transforming this problem from novelty detection into object detection.

The experiment was set up in parallel to what would hypothetically happen during deep space exploration. The training was done on a small batch of simulated images of ringed planets. The idea in the experiment is that, in theory, we have never seen a ringed planet before. Yet, our physics-based models of planet formation dictate that they would naturally occur. So we collect simulated images, train on that, and then send it deep into space. Upon finding a ringed planet for the first time, it would need to recognize those planets. Normally, we wouldn't be able to do this since we have no ringed planets to train on (in this hypothetical experiment), but since we used simulated images, we now have a model to deal with this. The experiment goes through this entire process, and even tests the model on real images of Saturn. Again, the machine has only seen a small batch of randomly generated simulated images of hypothetical ringed planets, never a real image of a planet. The results of the experiment are extremely positive and can be seen in the table below.

\vspace{5mm}

\begin{center}
 \begin{tabular}{||c c c||} 
 \hline
Type & Accuracy & Detection Errors \\ [0.5ex] 
 \hline\hline
 Planet & 99.22 & None \\ 
 \hline
 Rings & 99.11 & None \\
 \hline
\end{tabular}
\end{center}

\vspace{5mm}

As we can see, the model is dependably accurate based on solely simulated images. This experiment shows a key point of using simulators - by combining planet generation theory and realistic rendering, we have turned a novelty detection problem into an object detection problem, which is significantly easier to deal with. Now, instead of having to detect unknown features, we can simply construct planets randomly based on physical laws and train a model using those simulator images. 


\section{Simulator for Novelty Ranking}

We have shown previously that simulator images can be used with astounding accuracy, and with mass production, can make training via real images unnecessary. Therefore, we can train using hundreds or thousands of simulated images and when we encounter a planet, we can detect it, image it, and send those images back to Earth.

Say, for instance, that the wafer passes and images the five planets in a hypothetical solar system. Soon after that, it may be on an inevitable course toward that solar system's star, which will destroy the wafer and all of the images. One downside to small wafers is that they are easily destroyed or corrupted. This makes a priority system vitally important, as it would allow the wafer to possibly send back one or two images from the five that it collected before it is destroyed. This section is dedicated to figuring out which images should be sent back, and discussing the approach in doing so.

\subsection{The Concept}

We will assume that we have a small storage of images that we need to send back to Earth in an order that is based on importance. Wafers could be destroyed relatively easily and data transmission rates in space are very slow, so sending data based on a notion of importance is paramount.

\begin{figure}[ht!]
  \includegraphics[width=\linewidth]{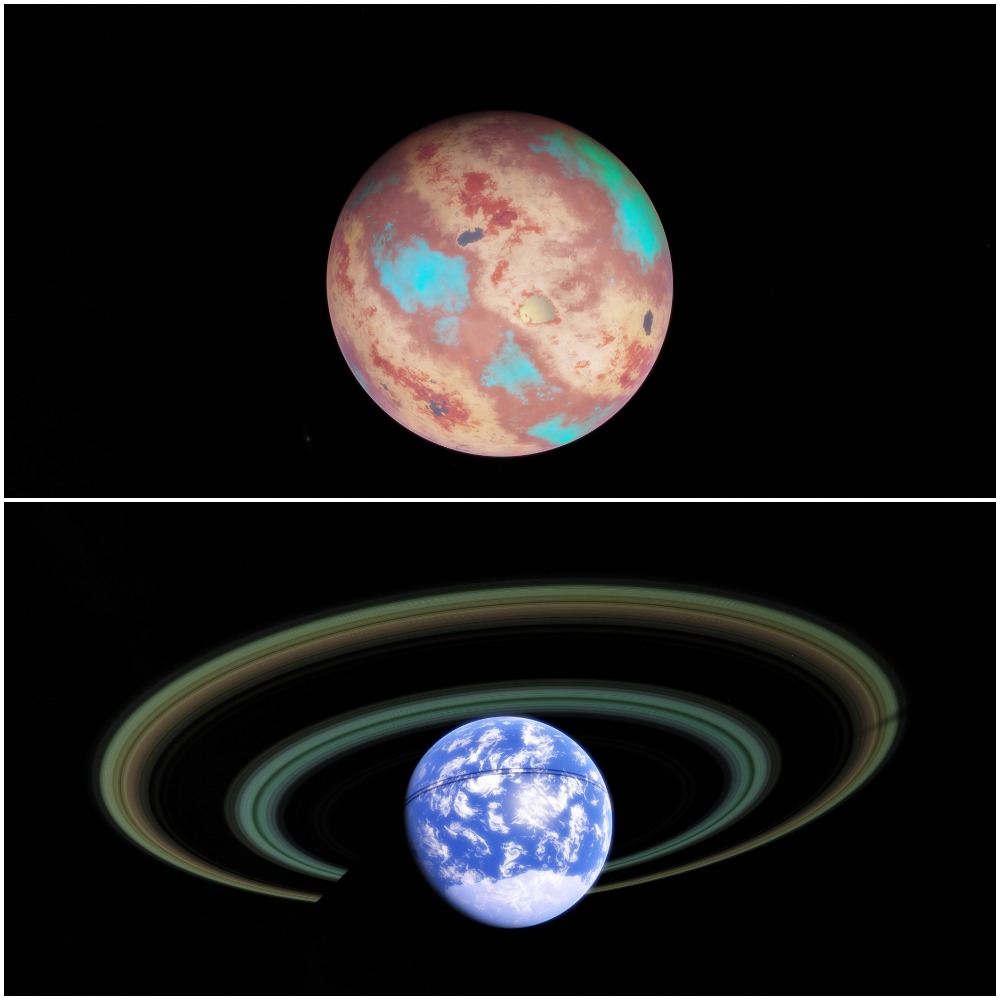}
  \caption{Hypothetical storage of two images that need to be ranked based on importance.}
  \label{fig:Comparison}
\end{figure}

Figure \ref{fig:Comparison} helps show us the extremely abstract definition of importance that we, as humans, may place on new planets. The top planet is colorful and full of land and different bodies of liquid, while the bottom planet has a unique double ring, an atmosphere, and a single large ocean, one that may be assumed to be water by visual inspection alone. The main question we want to ask here is: If you could only send one of these images back to humanity, which would you send?

An astrobiologist might choose the top planet since the presence of land and many differently-composed bodies of liquid exist, giving multiple opportunities for life to possibly flourish. Yet, someone interested in another planet that may be able to accommodate human existence might choose the bottom planet since it seems to offer two important features for us, water and an atmosphere. The question of importance to humans is very subjective, yet we need a solution that would be able to rank these two planets, and many more, in order of importance.

\subsection{The Human Experiment}

The implausibility of teaching vast conceptual knowledge to a machine in hopes of it gaining context made us seek out a different approach:

\begin{enumerate}
    \item Generate simulated images of planets that range in features. This will remove the bias that some people may have about our own solar system, since we are not using any real images of our own planets. It will also allow data to be gathered about features that do not currently exist in our solar system, but based on astrophysical theory, could exist in the universe. 
    \item Ask experimental subjects to rate each planet by \textit{Importance}.  This is posed via the question: "On a scale from 1-7, how important would it be for humankind to see this image if it were gathered by a spacecraft during deep space exploration?"
    \item Ask experimental subjects to rate each planetary feature. This will comprise our total planetary feature set. For instance: On a scale from 1-7, how much does this planet exhibit the presence.. of rings? ..of an atmosphere? ..of moons? ..of a livable environment for humans? 
    \item Using the data gathered from human thought processes and individual analysis of importance and interestingness of a planet, train a model to predict importance given a feature set.
    \item Rank all planets in storage based on importance and send them back to Earth via this priority system.
\end{enumerate}

This process takes in human definitions and thought processes in order to break down the concept of what we find interesting in planets that we have not even seen before. Using this method, we can bypass a problem of novelty detection, which is difficult, and machine contextual learning, which is extremely difficult, and turn it into a problem of human information gain, which is easy, and object detection, which is also easy.

Future work will show experimental results beyond this proof of concept solution.


\section{Simulator for Energy Management}

During a deep space voyage, the wafersat will need to be supplied with enough energy to perform necessary functions, such as imaging, analyzing the images, and transmitting data. We don't assume that the system is perfect, nor do we need certain restrictions on the amount of energy available. We have one goal: minimizing the amount of energy needed while ensuring planetary detection. At one end of the spectrum is full energy conservation, which would mean that the camera never turns on and therefore we never collect any data. On the other end of the spectrum is full energy use, meaning that the camera never turns off until the energy runs out, which would yield us many images but most likely none with important findings. Somewhere in between is optimal, but how do we find it? 

\subsection{The Two Phases}
Simulators open a whole new universe that can be utilized in order to make a virtual interstellar journey to Alpha Centauri hundreds of times in the span of a day. By doing this process, we can train our models to identify stars, predict distances, swap between the two possible phases, and in doing so, save energy while capturing meaningful images.

\subsubsection{Phase One}

Phase One is essentially comprised of time spent in open-space travel. This would mean that the probe is beyond a 'fair' distance away from any nearby star and that planetary detection would be a fruitless endeavor. Yet, during this phase, the main objectives would be nearby star detection and star distance predictions. 

\subsubsection{Phase Two}

Phase Two would be a rare occurrence whereas the probe has traveled within a 'fair' distance of a star and we no longer need to deal with nearby stars until we have left that star's system. Instead, this phase would prioritize planetary detection, imaging, and ranking. 

\subsection{The Process}

The trip from Earth to Alpha Centauri can be done in approximately 20 years. But, in the simulator, one can travel at any speed and cut out the majority of the time spent in an uneventful space. This makes it possible to simulate a 20 year journey in a few hours, or many journeys in just a single day.

Once these are done, we can train a machine learning model using star type, the section of the image containing the star, and the distance from the probe to the star (a simulator feature). Combined with a subtraction algorithm, and only using enough energy to take two images, the machine will be able to identify stars and predict their distance from the probe. 

Using this information, the probe will know the approximate distance to the nearest star in its forward path. A simple calculation can tell it a safe amount of time to wait until it should take two more images, confident that the time it has waited has been uneventful. 

Repeating this process is extremely energy efficient, and should eventually lead to coming within a reasonably 'fair' distance from a star. When this occurs, we would change into Phase Two. 

Phase Two would use the same intuition except that instead of stars, we substitute in planets. Once identified, instead of being interested in distances, we would prioritize imaging. Details on planetary detection, imaging, feature extraction, and ranking have been detailed in earlier sections.

\subsection{An example of Phase One}

One extremely difficult concept in this entire process is making sure that the probe can successfully understand what is close to it versus what is very far away. The concept used is straight-forward: bodies that are closer will tend to shift more while the probe travels in a straight path. As an extreme example, a body that is 1 AU away from the probe will shift from center screen to completely off screen in approximately 33 seconds. Yet, a very distant star could go without changing position for months or years. 

In order to deal with this, a subtraction algorithm is implemented. The probe will take a photo, wait a certain amount of time, and take another photo. Then, the first will be subtracted from the second and the resulting image will show any pixels that have shifted state during the elapsed time. If enough of these pixels shift, we will get a clear image of something that is relatively close. 

The main problem here, again, is that nobody has any concept of the "wait a certain amount of time" part of the process. How much time is the right amount of time? If you do not wait long enough, nothing will move and your subtracted image will be all black. If you wait too long, even things that are very far away will begin to shift and you will be left with a large amount of stars, still unsure about which of those are actually close. This difficult part becomes approachable with the use of simulators. 

The example deals with a simulated star that exists 0.08 light-years away from the probe. We travel at 500c and perform a subtraction algorithm in 10 second intervals, resetting after each one. This equates to traveling at 0.25c and performing a subtraction algorithm every 20,000 seconds, or approximately every 5.55 hours. So, the first image is 5.55 hours in real-time, the second image is 11.11 hours in real-time, then 16.66 hours, and so on. The goal is to see if simulators can be useful, and if so, at what point we would want to optimally take images in order to ensure we capture bodies that are nearby while also saving energy. 

\begin{figure}[ht!]
  \includegraphics[scale=0.5]{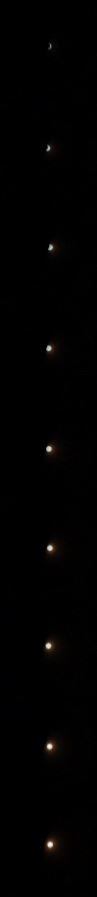}
  \caption{Subtraction algorithm performed in 20,000 second intervals}
  \label{fig:subtraction}
\end{figure}

As we can see in Figure \ref{fig:subtraction}, the first few bins produce a hazy image of the target star. At the fifth image, which would have the probe waiting approximately 28 hours between images, we can see a full image of the star. By the last image, which is represented by approximately 50 hours of real time, other nearby stars were showing very hazy signs of recognition from the subtraction algorithm. 

This proof of concept is extremely vital to star recognition and energy management. Depending how far away from a star we want the probe to be when it is able to recognize it, this process can be altered and honed easily.


\section{Conclusion}
We began with a set of new challenges that arise from the Starlight program and the ability to perform fast interstellar travel. These include identifying stars, identifying new planets, extracting never-before-seen features, conceptually ranking these new planets against each other in terms of importance, understanding what importance means in the context of planets, and conserving energy while performing needed tasks. 

We started off by showing that a simple classification model would not suffice. Not only does it perform poorly, but it does not come with the range of tools that are needed for further processes down the line.

We show that while training on simulated images, our accuracy on real images does not suffer, which is an astounding concept for such an application. Along with this, we provide results and many reasons why simulators enable us to identify features that we have yet to observe in actual images. With the use of simulators, we can run experiments on humans in order to extract the features of an important planet and use that knowledge to dictate decisions for planetary importance rankings. 

Lastly, we demonstrate how simulators can be utilized to save energy while ensuring that all necessary functions are completed. 

There is much planned future work on this topic. This includes optimizing simulators for this specific task, choosing the best hardware given restraints such as size and the harsh environment of space, and incorporating more human knowledge gain into the AI process.

\section{Acknowledgements}
%

PML gratefully acknowledges funding from NASA NIAC
NNX15AL91G and NASA NIAC NNX16AL32G for the NASA Starlight program
and the
NASA California Space Grant NASA NNX10AT93H,
a generous gift from the Emmett and Gladys
W. Technology Fund, as well as support
from the Breakthrough Foundation for its Breakthrough StarShot
program.
More details on the NASA Starlight program can be found
at \url{www.deepspace.ucsb.edu/Starlight}.

\section{Citations}
\label{Section 3}




\end{document}